\documentclass[12pt]{article}
\usepackage{amsmath}
\title{\bf The nature of electronic charge}
\author{{\bf S.C. Tiwari} \\ {\it Institute of Natural Philosophy} \\ {\it 1 Kusum Kutir, 
Mahamanapuri} \\ {\it Varanasi 221005, India} \\ {\it Email: vns\_sctiwari@yahoo.com}}
\begin{document}
\maketitle
{\small Advances in gauge theories and unified theories have not thrown light on the meaning of electron. 
The problem of the origin of electronic charge is made precise, new insights gained from Weyl space are 
summarized, and the origin of charge in terms of fractional spin is suggested. New perspective on the 
abelian Chern-Simons theory is presented to explain charge.}


\section{INTRODUCTION}

One of the key ideas on which modern unified theories of fundamental forces of nature are founded is that 
of the non-abelian gauge theory propounded by Yang and Mills in 1954 \cite{1}. However, historically the 
first non-abelian gauge theory for weak and electromagnetic interactions was due to Klein in 1938, 
see \cite{2}. This review by Jackson and Okun \cite{2} also draws attention to the little known fact that 
the covariant gauge condition in electrodynamics commonly referred to 
as  the Lorentz condition was first introduced by L.V. Lorenz. It may also be noted that 
electromagnetism is a U(1) gauge theory only under certain restrictions \cite{3}. Since the Standard 
Model (SM) of particle physics is a gauge theory with the gauge group SU(3) X SU(2) X U(1), 
and there are strong reasons to look beyond SM or seek 
alternatives \cite{4}, the preceding remarks are quite pertinent. It is well known that quantum 
electrodynamics (QED) is a paradigm for unified gauge field theories, and it explains empirical 
observations with remarkable degree of accuracy. However, Dirac one of the founders of QED remained skeptical 
about the renormalization procedure, and perhaps in his last criticism published in 1984 \cite{5} 
advocated drastic change in its foundations. In my monograph \cite{4} imprint of Dirac's views 
is obvious, however the critique offered by Dyson \cite{6} is also worth 
mentioning. In QED a physical quantity is represented by a perturbative power series expansion 
in the power of the fine structure constant. Renormalization of charge and mass ensures that the 
series is term by term finite i.e. each coefficient in the series is finite. Dyson asks 
whether the convergence of the series as a whole could be proved. Introducing a fictitious world 
in which like charges attract each other, and exploiting the 
mathematical property of the power series, he shows that the series would be divergent. He favors the implication 
that QED may not be a closed theory on mathematical grounds, and that it is a half-theory admitting the 
possibility of new ideas. I think this is a reasonable assessment. 

For a constructive alternative we add to this the question: Why renormalization succeeds? In fact there is an 
important result related with this that concerns the role of chiral anomaly \cite{7}. It is known that 
massless QED Lagrangian is invariant under chiral phase transformation, and the conservation of the 
corresponding Noether axial-vector current is true only to the tree approximation in the 
perturbation calculation. For axial-vector and two vector vertices there is an anomalous term; this has come 
to be known as Adler-Bell-Jackiw anomaly or chiral anomaly. In QED it is not possible to eliminate it 
without violating gauge invariance, renormalizability or unitarity \cite{7}. Surprisingly in the 
electroweak SM the condition for anomaly cancellation is satisfied; note that 
photon does not couple to axial-vector current but weak gauge bosons couple to chiral currents. Is there a 
deep reason for this result?  I believe a fundamental question has remained practically unnoticed: 
What is the meaning of charge? In the monograph \cite{4} 
we raised this question, and suggested that a viable alternative to the current approach for unification 
would have to answer this question. Associating charge with internal gauge symmetry is a well known 
prescription; therefore our question has to be precisely defined. Historically beginning with the 
attempts of J. J. Thomson, the problem of charge has been investigated relating the 
electromagnetic energy with the inertial mass of the electron. Equation of motion of a radiating electron, 
spinning extended model of the electron, and role of gravitational field in the classical electron model 
have been widely discussed, and reviewed in \cite{4}. 

The nature of charge is not a subject in all these works. Dirac in 1931 postulated a new charge, 
namely the magnetic monopole, and argued that consistency with quantum theory for a monopole and electronic 
charge system would lead to the quantization of charge \cite{8}. Note that even the presence of one 
monopole in the universe would imply that charge is an integral multiple of a 
smallest unit of charge. However Dirac's hypothesis does not explain what charge is. In this paper we 
propose a radically new idea: electronic charge is a manifestation of fractional spin angular momentum. 

The aim of the present paper is to make precise the problem of electronic charge, delineate the new 
insights gained from the Weyl geometry \cite{9}, and explore the Chern-Simons field theory to understand 
the meaning of charge in terms of fractional spin angular momentum. In the next section a brief review 
of the notion of charge is presented and the problem of the meaning of charge is elaborated. 
Three interpretations of monopole physics are discussed to argue that the electron charge problem 
remains unsolved. In Sec.3 important implications of Weyl space for the electron model are discussed. 
Is it possible to picture charge itself in terms of gauge field in 2+1 dimensional space? Sec.4 explores 
Chern-Simons theory in this context in order to interpret charge as a manifestation of fractional 
angular momentum. Philosophical implication of the Yang-Mills theory on the source-field duality 
is in consonance with the approach presented here. Recent reports though inconclusive suggest that the 
fine structure constant may be varying with the age of the universe. These aspects are commented upon 
in the concluding section.

\section{STATEMENT OF THE PROBLEM}

Though Faraday anticipated atomicity of electricity from the phenomenon of electrolysis, it was the 
clear statement 
of atom of electricity by Helmholtz in 1881 that led to an intense debate in the context of aetherial 
world-view in the late 19$^{th}$ century. In the last century, quantized charge, electromagnetic model of 
electron, and the problem of intrinsic  spin have received a great deal of attention; in fact, even 
after the advent of SM such classical problems have continued to arouse curiosity, we refer to a 
comprehensive review in \cite{4}. The physical meaning of electronic charge is not clear in 
spite of the advances in theoretical physics; unfortunately, that there exists a problem does not 
seem to be recognized. Let us elucidate the problem in the following.

In the Maxwell-Lorentz electrodynamics, the charge and current density represent the sources for the 
electromagnetic fields. The current continuity equation embodies the charge conservation law. 
In the 4-vector relativistic notation, the current continuity equation is given by 
\begin{equation}
\partial_{\mu}J^{\mu} = 0 \; ,
\end{equation}
and corresponding charge is defined to be
\begin{equation}
Q = \int J^0 dV \; .
\end{equation}
The charge $Q$ is conserved if $\bf J$ satisfies suitable asymptotic boundary conditions.

The famous Millikan's oil drop experiment was considered the best measurement of the electronic charge, 
the name electron for the postulated fundamental unit of electric charge was coined by Johnstone 
Stoney.  For a point electron, using Dirac delta function we can write the expression for charge density as
\begin{equation}
\rho_e = e\delta({\bf r}) \; .
\end{equation}

Macroscopic charge distribution in the continuum limit could be represented by a smooth function, $\rho$ and 
the differential equation for electric field in vacuum is given by
\begin{equation}
\nabla \cdot {\bf E} = 4\pi \rho \; .
\end{equation}

The numerical value of electronic charge depends on the system of units one chooses, and the standard 
theory does not attach any fundamental significance to the choice whether current (or charge) is a 
fourth basic dimension other than those of length, time and mass or whether the electromagnetic 
quantities are expressed in terms of the three basic mechanical dimensions \cite{10}.  In quantum 
field theories, a system of natural units is convenient such that $\hbar = c= 1$. Action function is 
dimensionless in natural units, and all physical quantities are expressed in the units of power 
of mass. For example, the length dimension is (mass)$^{-1}$. Note that all physical measurements 
ultimately reduce to the counting of the measurement of calibrated spatial relations; therefore, we 
argue that the length dimension should have fundamental significance. In the case of electron, 
the Compton wavelength, $\lambda_c$ and the electron charge radius, $a_e$ are two characteristic 
lengths. Interestingly, as is known their ratio is equal to the electromagnetic 
coupling constant, $\alpha$.

A physical explanation is usually sought in terms of gauge symmetry and Noether theorem: invariance 
of Lagrangian under U(1) gauge transformation leads to a conservation law via the continuity equation, 
e.g. Eq. (1). In QED, electron and electromagnetic field interaction is obtained using a minimal 
coupling prescription
\begin{equation}
\partial^{\mu} \to \partial^{\mu} + ieA^{\mu} \; ,
\end{equation}
 in the Lagrangian of free Dirac electron of mass $m$
\begin{equation}
L_0 = \psi^{\dagger}(i\gamma^{\mu}\partial_{\mu} - m)\psi \; .
\end{equation}

Local gauge transformation of $\psi$  leaves the total Lagrangian invariant if $A_{\mu}$ also 
undergoes the gauge transformation, namely
\begin{equation}
A^{\mu} \to A^{\mu} + \partial^{\mu}\alpha \; .
\end{equation}

Gauge invariance leads to a local conservation law similar to Eq.(1). In quantum theory the charge 
is a Hermitian operator  $\hat Q$ and it is necessary that quantum vacuum state is also invariant 
under U(1) transformation which implies
\begin{equation}
{\hat Q}\mid 0 > = 0 \; , 
\end{equation}

Note that in the SM, spontaneous symmetry breaking where the above condition is not satisfied 
plays an important role. Since the internal space of U(1) is an artifact unrelated with the space-time, 
this sort of explanation does not throw any light on the physical nature of charge.

There is an interesting argument that provides the reason for the quantization of charge 
postulating new particle with a magnetic charge, $g$ i.e. the magnetic monopole. In the original 
formulation due to Dirac the vector potential is singular along a string 
extending from zero to infinity. Quantum mechanics of charge-monopole system leads to the quantization condition
\begin{equation}
\frac{eg}{\hbar c} = \frac{n}{2} \; .
\end{equation}

The monopole problem can be recast in terms of the Hopf mapping of  3-sphere to 2-sphere  such 
that the vector potential is non-singular defined on the local sections \cite{11}.  
The third important scenario of monopole physics is that related with the 
rotational invariance and angular momentum. Saha-Wilson quantization is derived based on the 
angular momentum quantization \cite{12}. In a nice paper, Goldhaber investigates the significance 
of spin in the charge-monopole scattering problem \cite{13}.He considers scattering of a spinless 
charge from a spinless monopole, and applies quantum rules. It is found that an extra 
spin arises, and the Dirac quantization condition also emerges. This intrinsic spin cannot be 
ascribed to charge or monopole independently, but to both. Instead of charges 
(electric and magnetic) as elementary objects, I think this discussion shows that
angular momentum which is a mechanical quantity should be treated as fundamental. 
We carry the argument further in what follows.

If we examine the Maxwell field equations then the unit of charge can be factored out leaving 
the electromagnetic quantities possessing purely geometric dimensions for example, 
{\bf E} in Eq. (4) would have the dimension $L^{-2}$  ($L$ is length dimension). 
On the other hand, the Lorentz force acquires a multiplication factor of $e^2/c$, and this 
factor also accompanies observable quantities like electromagnetic momentum and energy. 
Interestingly the dimension of $e^2/c$ is that of angular momentum. 
These considerations suggest that the origin of electronic charge may be mechanical. 

The fine structure constant can be interpreted as a ratio of two angular momenta
\begin{equation}
\alpha = \left(\frac{e^2}{c}\right)/\left(\frac{h}{2\pi}\right) \; .
\end{equation}

An important property of electron is its magnetic moment, and it is now established that 
electron magnetic moment to first order 
in $\alpha$ is given by
\begin{equation}
\mu_e = \mu_B\left[1 + \frac{\alpha}{2\pi}\right] \; .
\end{equation}

Here the Bohr magneton $\mu_B = eh/4\pi mc$. Eq. (11) can be re-written as
\begin{equation}
\mu_e = \frac{e}{mc}\left[ \frac{h}{4\pi} + \frac{e^2}{4\pi}\right ] \; .
\end{equation}

For historical reasons the first term in the square-bracket in Eq. (12) was identified with 
the spin-half of the electron since the precision of experiments was not adequate to infer 
the anomalous term. We suggest that electron spin has fractional part too, and 
that it is intimately related with the origin of charge \cite{4}.

To a critic for attaching deep significance to $e^2/c$, we remind that the first decisive 
step to unify electricity, magnetism and optics was based on a single empirical fact: 
the near equality of the velocity of propagation of electric and magnetic field disturbances in a 
dielectric and the velocity of light. As regards to the significance of this problem, 
we quote Kiehn \cite{3}: ``The origin of charge has long been a mystery to physical 
theory, perhaps even more elusive than the concept of inertial mass", and refer to 
Weyls speculation on this question \cite{9}. Puzzling aspects of charge in supergravity 
and brane theories have been recently noted in the literature; 
a nice discussion is given by Marolf \cite{14}.

\section{NEW INSIGHTS FROM WEYL GEOMETRY}

Weyl's was the first unified theory of electromagnetism and gravitation \cite{9}, 
and though Weyl himself abandoned it, the beauty and philosophical motivation behind 
it (the transcendental logic) make it still attractive. We do not go into the formal aspects here, and 
refer to \cite{9}, and detailed discussion in \cite{4} for this purpose. An important 
feature of Weyl's theory is that the charge-current density, $J_{\mu}$ is proportional 
to the electromagnetic four-vector potential $A_{\mu}$ (note that in Weyl's theory the field 
quantities have geometrical dimension). This implies that the field acts as a source for 
itself, and according to Weyl, electric charge and current are diffused thinly 
throughout the world. A similarity with the Yang-Mills theory is obvious: the gauge 
fields carry charge. Weyl admits that the inequality of positive and negative electricity 
is not explained in his theory, and indicates that it may be linked 
with `unique direction of progress characteristic of time, namely past $\to$ future'.

A promising new approach to Weyl's geometry is based on a gauge covariant bimetric 
tetrad space time \cite{15}. Israelit in this paper \cite{15} while writing the action 
integrals includes the action. $I_m$ for the matter, and obtains source-density four-vector for 
the electromagnetic fields comprising of two parts: $J_{\mu}$ and $V_{\mu}$. Curiously the 
part $V_{\mu}$arises due to the spin angular momentum of matter, and the author notes that 
if conventional charge-current density is absent, ``the spin angular momentum 
density tensor of matter will induce a nonzero `current density' vector.....". 
For recent account of this work, see the monograph \cite{16}.

In our model of electron, we envisaged 2+1 dimensional internal structure, endowed physical 
significance for characteristic lengths associated with the electron, namely the Compton 
wavelength and charge radius, and visualized it in terms  of bound fields 
($f, g$) and a circulating field ${\bar f}$ accounting for charge \cite{17}. In 1989, 
the Weyl geometry was explored to formulate this model \cite{18}. Recently we have 
investigated the Einstein-Weyl space to model the electron \cite{19}. Though the progress
in this approach is not conclusive, new insights have been obtained. In the original 
Weyls theory there does not exist a scalar corresponding to the scalar curvature 
though there does exist a distance curvature. In \cite{18} an in-invariant scalar, $\psi$ was 
introduced to seek interpretation of charge, constructing a co-scalar, $\xi$, 
\begin{equation}
\xi = g^{\mu\nu} \partial_{\mu} \psi \partial_{\nu} \psi \; .
\end{equation}

This formulation indicates that the sign of charge is determined by time reversal transformation. 
However, there was no explanation for spin. A formal approach is to incorporate spinor field 
in the Lagrangian, but it really does not offer a solution to the problem. An insight 
to the hidden structure of $\psi$ is found if we argue that $\sqrt{\xi}$ is linearly related with 
$\psi$. A nontrivial but interesting possibility is that
\begin{equation}
\sqrt{\xi} = i \gamma^{\mu} \partial_{\mu} \psi \; .
\end{equation}

If we impose the condition that the field is decoupled then this equation gives the Dirac equation 
for massless spinor. If we reflect on the phenomenological model of electron \cite{17}, and learn 
from Weyl's theory generalized in \cite{18} then it becomes clear that the explanation of 
electronic charge would emanate from the gauge field, somehow representing the field $\bar f$, 
and the sign of charge would be determined by time reversal symmetry: in the next section we 
consider Chern-Simons theory towards this aim.

To summarize: source-field duality seems to be an impediment in the unification, and Weyl's 
geometry hints at the possibility of dissolving this separateness; that gauge fields may 
originate charges is an idea in consonance with Yang-Mills theory as well as 
supergravity - membrane theories \cite{14}, and the role of time-symmetry and spin angular 
momentum for understanding charge is an exciting possibility.

\section{ORIGIN OF CHARGE - CHERN-SIMONS FIELD THEORY}

The presence of Chern-Simons (C-S) term modifies the Bianchi identities, and a natural question 
arises whether the gauge invariant fields could be interpreted as the sources. Marolf \cite{14} 
has defined and elucidated three notions of charge in supergravity and 
brane theory inspired formulations : brane source charge, Maxwell charge, and Page charge. 
The Maxwell charge is conserved, but not localized. We have discussed in the preceding 
section that in the Weyl theory charge is diffused in the whole space, and in
the Yang-Mills theory gauge fields possess charge, therefore seeking origin of electronic 
charge in terms of the abelian C-S theory seems reasonable. Let us make it clear that the 
internal structure of electron is visualized in a plane from which a point is 
removed i.e. a punctured plane, and the internal fields have phase velocity equal to the 
velocity of light, therefore unlike usual rigid extended models there is no conflict with 
relativity. The electron motion in three dimensional space is such that the time 
periodicity in core region is in synchronization with the motion along, say z-direction.

Physics in (2+1) dimensional space time has many interesting and intriguing aspects; however 
our discussion is limited to the abelian C-S theory. We refer to \cite{20} for a 
self-contained description of C-S theories, and \cite{21} for Deser's nice 
survey of 2+1 D theories. The standard Maxwell plus C-S Lagrangian for the abelian case is given by
\begin{subequations}
\begin{eqnarray}
L &=& L_M + L_{CS} \; ,  \\ 
L_M &=& -\frac{1}{4p}F_{\mu\nu}F^{\mu\nu} \; , \\
L_{CS} &=& \frac{\kappa}{2} \epsilon^{\mu\nu\sigma}A_{\mu}\partial_{\nu} A_{\sigma} \; .
\end{eqnarray}
\end{subequations}

Note that $L_{CS}$  is not manifestly gauge - invariant, however under the gauge transformation 
a total space-time derivative arises that vanishes at the infinity (neglecting boundary terms). 
Variational principle leads to the equation of motion
\begin{equation}
\partial_{\mu} F^{\mu\nu} = -\frac{\kappa p}{2} \epsilon^{\nu\rho\sigma} F_{\rho\sigma} \; .
\end{equation}

The charge-current density for matter as an external term interacting with the field is 
introduced in the form $A_{\mu}J^{\mu}$ . For a simple case of only C-S term, the Gauss 
law assumes the form
\begin{equation}
\kappa B = \rho \; ,
\end{equation}

while for Maxwell + C-S theory, we have
\begin{equation}
\nabla \cdot {\bf E} + \mu {\bf B} = \rho \; .
\end{equation}

Here $\mu = p \kappa$. In the standard interpretation the structure of Eq.(16) suggests a 
topologically massive gauge theory, and Eq.(17) implies that charge and magnetic field are 
linked. In fact, due to finite range behaviour of {\bf E}, Eq. (18) when integrated over all space gives
\begin{equation}
\mu \Phi = Q \; ,
\end{equation}
where magnetic flux, $ \Phi = \int B d^2 x$ and $Q = \int \rho d^2 x$. A physically motivated 
illustration is to consider an electron confined in a 2D plane, and a magnetic field is 
applied normal to the plane. This electron-flux tube composite is described by the C-S 
theory. This approach has found many fruitful applications, specially in (fractional) quantum Hall 
effect \cite{22}. 

We make a radical departure, and ask whether the electron itself could be considered a 
charge-flux composite. Note that in the standard interpretation, electron is assumed a point 
charge circulating around an externally applied magnetic flux tube; obviously one neglects 
electron magnetic moment i.e. the charged particle could more appropriately be considered 
to be charged pion. Instead of this, let us imagine electron to be an extended 2+1 D structure  
such that the short range fields determine the core, the magnetic field B is the self 
field, and Eq. (18) with $\rho = 0$  defines charge in terms of the flux
\begin{equation}
\mu B = - \nabla \cdot {\bf E} = - \rho_e \; .
\end{equation}

Asymptotically the fields $A_i$ are pure gauge, thus the charge originates due to pure 
gauge fields. Since the C-S term changes sign under time reversal, the sign of charge is 
related with time reversal transformation. The pure gauge field extends to infinity, 
thus the charge in a sense, is distributed over whole space. It is significant that C-S 
term does not contribute to the energy-momentum tensor, and the total energy, for example, 
is due to Maxwell term that dominates in a small region which we term the core of the 
electron. Is core dimension characterized by Compton wavelength of the electron? 
The complete solution of the field Eq. (16) is being investigated with this new 
perspective. If we express magnetic field in the usual 4D theory in the unit divided 
by $e$, then the basic flux quantum $hc/e$ becomes $hc/e^2$.  This value is same as the ratio 
of two terms in the magnetic moment of electron, namely $h/4\pi$ and $e^2/4\pi c$. 
This is suggestive of constructing a model of electron that could explain charge and 
spin in terms of 2+1 D fields. Note that the C-S term gives a topological invariant, 
therefore it seems to account for the quantization of charge.

\section{CONCLUSION}

Plausible arguments are presented to seek origin of electronic charge in terms of pure gauge 
field in the C-S theory linking fractional angular momentum with charge and time inversion 
determining the sign of charge. Developing a model of electron based on this idea seems an 
interesting possibility though a very difficult one. We guess that the core region fields could lead 
to a half-quantized vortex. Note that the study of half-quantized vortices in a different 
context is an active field of research \cite{23}. The essential point is that for purely 
space-time model of elementary particles a new approach is required. Our model of electron 
and interpretation of charge in terms of internal rotation deserves attention towards this aim. 

Recent reports indicate that the fine structure constant may not be a constant and may have 
temporal variation. Most stringent laboratory limit is obtained using cesium atomic clocks 
\cite{24}. Quasar absorption spectra provide astrophysical limits \cite{25}. 
In my book on superluminality \cite{26} variation of fundamental constants is reviewed. 
In \cite{26} we argue that time varying velocity of light is natural, and also that it 
is in harmony with the envisaged space-time model of physical objects. Finally we 
mention that if the idea that charge has origin in rotation succeeds it would be in 
conformity with our reinterpretation of Maxwell electrodynamics as photon fluid \cite{27}.

An important question raised by one of the referees concerns the time variation of charge 
in case the temporal variation of the fine structure constant is established. It is 
pertinent to discuss certain observations made by Dirac in this context. He argues that 
since the inverse of the fine structure constant is close to a prime number, not all the 
three quantities comprising it could be fundamental. Vacuum velocity of light plays such 
an important role in relativity that it has to be fundamental. Further, if the 
charge is not fundamental it will appear as square root in the basic equations-not a 
desirable thing. Therefore Dirac suggests that the Planck constant may not be fundamental. 
Now time-variation of ? introduces another possibility since all the three
quantities could depend on time. In \cite{26} different possibilities discussed in the 
literature have been reviewed.  In particular the approach due to Bekenstein \cite{28} 
has been revived in recent years. The variable charge model has the virtue that it could be 
used to account for time-varying fine structure constant without making any radical departure 
from the basic principles, e.g. Lorentz invariance, gauge invariance, and general 
covariance. Nielsen-Olesen vortex-like solutions and cosmological implications 
of variable charge models have been briefly reviewed in \cite{26}. Recent paper by 
Bento et al \cite{29} follows the Bekenstein approach, and considers quintessence 
fields coupled to the electromagnetic field. Besides the cosmological consequences, 
the limits from the empirical data are also examined. I think the authors make a 
reasonable conclusion that so far time-varying $\alpha$ lacks firm observational evidence. 
In our interpretation of charge, the ratio of two angular momenta determines the fine structure 
constant, and for an evolving universe it would be natural to expect that it varies 
with time, however as noted above we prefer time-varying velocity of light at a basic level.

{\bf Acknowledgements}

I am grateful to Professor M. Israelit for drawing my attention to his contributions 
on Weyl theory. I thank the referees for encouraging comments that led to the improved 
presentation. The library facility of Banaras Hindu University, Varanasi is acknowledged.

\end{document}